\documentclass[11pt,twoside]{article}
\usepackage{newpasp}
\def\edcomment#1{\iffalse\marginpar{\raggedright\sl#1\/}\else\relax\fi}
\reversemarginpar

\begin{document}
\title{On the Origin of the Fanaroff--Riley Dichotomy}
\author{Gopal-Krishna}
\affil{National Centre for Radio Astrophysics, TIFR, Pune University
Campus, Post Bag No.\ 3, Ganeshkhind, Pune 411007, India}
\author{Paul J.\ Wiita}
\affil{Princeton University, Department of Astrophysical Sciences,
Princeton NJ 08544-1001, USA; on leave from Georgia State University}

\begin{abstract}
A small fraction of double radio sources show a peculiar and
striking hybrid morphology; they have a distinctly FR I  structure on one 
side of the nucleus, and a FR II structure on the other.  We argue 
that the mere existence of these HYMORS is quite incompatible with 
the theoretical explanations for the Fanaroff--Riley dichotomy that 
are based upon the nature of the jet plasma, or those invoking 
an intrinsic property of the central engine.  Rather, these HYMORS 
strongly support models that explain the difference between FR I 
and FR II sources in terms of asymmetry of interaction of the jets 
with the external environments.  We further show that a model for
radio source dynamics we had earlier proposed can neatly reproduce 
the observed dependence of the radio power dividing the two FR 
classes on the optical luminosity of the host galaxy, as found 
by Owen \& White and Ledlow \& Owen. 
\end{abstract}

\section{Introduction}
The vast majority of extragalactic radio sources can be easily classified
into the more powerful, edge-brightened FR II sources and the weaker, edge-dimmed
FR I sources (Fanaroff \& Riley 1974).  The boundary between these classes
in terms of monochromatic radio power was shown to rise with increasing optical
luminosity of the host galaxy: $P^*_R \propto L_{\rm opt}^{1.65}$
(e.g., Ledlow \& Owen 1996, Fig.\ 1). 

The large number of explanations for the Fanaroff-Riley dichotomy can be
divided into {\it intrinsic} models, where some property of the central engine
launches entirely different types of jets, and {\it extrinsic} models, where
very similar jets are ejected in both FR classes but interactions with 
the environment determine their large-scale morphologies (Gopal-Krishna \& 
Wiita 2000a --- GKW00,
and references therein).  Among the better developed intrinsic models are
those where: faster spinning black holes launch FR II
jets and FR I jets come from slower spinning BHs; jet plasma 
is composed of $e^-$--$p$ for 
FR II's  and $e^-$--$e^+$ for FR I's; advection dominated accretion disks yield
FR I's while standard thin disks yield FR II's. Possible 
support for such models comes from the differences between the properties of
the host galaxies of the two FR classes (e.g., Baum et al.\ 1995).

Various extrinsic models all argue that the jets
differ, if at all, only in their power or thrust, so the morphological
dichotomy arises from the differences in the interactions of the jets with the 
gaseous media. Weaker and/or poorly collimated jets, or those ploughing 
through a more disruptive environment, quickly lose their terminal
hot-spots (Gopal-Krishna \& Wiita 1988), 
or are subject to instabilities that dramatically increase the 
entrainment of ambient gas (e.g., De Young 1993; Bicknell 1984, 1995), both 
processes leading to a FR I morphology. Probably the strongest evidence
for the basic similarity of the jets comes from the VLBI observations
showing that the jets in most FR I, as well as FR II, radio
sources start off with 
relativistic bulk velocities; however, only FR II jets are able to 
remain relativistic to very large distances (e.g., Laing et al.\ 1999 
and references therein).

\section{HYMORS Support Extrinsic Models}

An observational clue for discriminating between the intrinsic and extrinsic
models comes from HYbrid Morphology Radio Sources, or HYMORS (GKW00).
These sources have a clearly FR I structure on one side of the host 
galaxy and a FR II structure on the other side. Some good examples 
gleaned from over 1000 published radio maps are listed in Table 1
(see GKW00 for maps and references). It is
seen that HYMORS occur among all basic radio loud AGN types ---
QSR, BL Lac and radio galaxy (RG); their linear sizes range from 
subgalactic to `giant'; their powers are close to $P^*_R$.  A few 
candidate HYMORS at higher redshift also exist and have 
powers above 
the nominal FR I/FR II break (GKW00).

HYMORS are extremely difficult to reconcile with radio jet models that
rely upon intrinsic differences in the central engines to produce the FR
dichotomy.  This entire class of models would predict that {\it both}
lobes of any double radio source are either FR I type or FR II type 
(rather than having distinctly different FR morphologies, as seen in
HYMORS).  On the other hand, within extrinsic models, modest differences 
in the clumpiness, density or velocity of the ambient medium on the two 
sides of the host galaxy could trigger the transition from a FR II jet 
to a slower jet producing a FR I appearance on one side (Gopal-Krishna
\& Wiita 1988; GKW00).  The more asymmetrical gas distributions 
suspected in HYMORS may be detectable, e.g., with {\it Chandra}.
 
\begin{table}
\caption{Properties of the HYMORS}
\begin{tabular}{cccccc}

\tableline
 Object &  Type & $z$ & Size & Size$^a$ & Log (L$_R$) \\
 & & &(arcmin)&(kpc) &(1 GHz) W/Hz$^a$ \\
\tableline

0131$-$367 & RG&0.029 &   14.2 & 483 & 25.4 \\
0521$-$364 & BL Lac&0.055 &  0.3   & 22  & 25.4\\
1004$+$130 & QSR&0.240 & 1.8    & 524 & 26.3\\
1452$-$517 & RG& 0.08 & 20.3   & 812 & 25.4\\
1726$-$038 & RG &(0.05)$^b$&0.6    & 35  & 24.8\\
2007$+$777 & BL Lac&0.342 & 0.5    &  213 &  24.8 \\
\tableline
\tableline
\end{tabular}

$^a$$H_0$ = $75$ km s$^{-1}$ Mpc$^{-1}$,  
$q_0 = 0.5$, spectral index = $-1$.

$^b$Estimated redshift from POSS plate; see GKW00.
\end{table}

\section{The Radio Power Division and Host Galaxy Luminosity}

The realization that only very powerful radio sources yield FR II
morphologies if their host galaxies are highly luminous (e.g., Owen \& 
White 1991; Ledlow \& Owen 1996) begged for an explanation, and several 
were put forward.  Although the `magnetic switch' model (Meier 1999)
can provide a decent fit to this\break $P_R^* \stackrel{\sim}{\propto} 
L_{\rm opt}^{1.7}$, relation, as an `intrinsic' model it is difficult 
to reconcile with the existence of HYMORS.  Venturing well outside the 
standard jet paradigm, the gravitational slingshot mechanism does 
appear able to both produce HYMORS and roughly reproduce the Owen--Ledlow 
relation (Valtonen \& Hein\"am\"aki 2000).  Considering the growth of 
instabilities in relativistic jets in an `extrinsic' jet model,
Bicknell (1995) put forward a complex, but plausible, explanation for 
this $P_R^* - L_{\rm opt}$ relation, though this model yields a slope 
of $\sim 2.1$. 

Here, motivated by our earlier study of `Weak Headed Quasars' (WHQs) 
(Gopal-Krishna, Wiita \& Hooda 1996), we propose a variant `extrinsic' scheme. 
We argued that that the lack of a hotspot in WHQs could be best 
explained through the onset of a jet's decollimation when the hotspot's 
(or, nearly equivalently, the bow shock's) velocity becomes transonic 
relative to the external medium and, therefore, the jet begins to
advance like a plume. The concomitant cessation of the `backflow' of 
the jet plasma accelerates entrainment into the jet, owing to the 
diminution of the protective cocoon around the jet which had hitherto 
separated the latter from the ISM material. This is further compounded
by the weakening of the confining effect of internal reconfinement 
shocks in the jet engendered by impinging cocoon vortices (e.g., 
Hooda \& Wiita 1998). These effects naturally produce FR I morphologies.

We follow Bicknell (1995)
by making use of the same empirical relations between
the elliptical's blue magnitude, $M_B$, and its soft X-ray emission,
$L_{\rm X}$, stellar velocity dispersion, $\sigma$,
and X-ray core radius, $a$.   However, we adopt a jet propagation
model which gives the hot-spot velocity, $v$, as
a function of distance, $D$, from the central engine 
(Gopal-Krishna, Wiita, \& Saripalli 1989):
\begin{equation}
v(D) = {\frac {X c[1+(D/a)^2]^{\delta/2}} {D + X[1+(D/a)^2]^{\delta/2}}}.
\end{equation}
Here $X = (4 L_b/\pi c^3 \theta^2 n_o m_p \mu)^{1/2}$,
where $L_b$ is the jet (beam) power, $\theta$ is the
jet's effective opening angle, $m_p$ and $\mu$ have their
usual definitions,
and $n_0$ is the central density of the ISM, which is
taken to fall off as (e.g., Canizares, Fabbiano, \& Trinchieri 
1987), 
$n(D) = {n_0}{[1+(D/a)^2]^{-\delta}}$.  Observations give
 $\delta \simeq 0.75$ and $\theta \simeq 0.1$ rad; also,
$n_0 \la 1~{\rm cm}^{-3}$ (e.g., Conway 2000).  

Noting that the transition to FR I morphology is typically seen to
occur at or before
$D = 10$ kpc, we equate $v(D^*) = c_s$ at $D^* \le 10$ kpc.
Clearly $c_s$ is related to $\sigma$, and empirically $\sigma$ depends 
directly on $M_B$, as does
$a$; furthermore $n_0$ and $a (\sim 1$ kpc) 
are related to $L_X$, which also depends on $M_B$.
We derive and solve the following equation for $L_b$ (via $X$) as a function of
$M_B$ 
\begin{equation}
{\rm dex}(-4.720 -0.0959 M_{\rm B}) = 
{\frac {X} {D^*/[1+(D^*/a)^2]^{\delta/2} + X}}.
\end{equation}
For $D^* = 10$ kpc, Eqn.\ (2) yields an approximate power-law: $L_b^* \stackrel{\sim}{\propto}
L_{\rm opt}^{1.56}$, normalized with $L_b^* = 3.6 \times 
10^{42}$ erg s$^{-1}$
at $M_B = -21.0$.  This produces an excellent fit to the slope of
the FR division in the Owen--Ledlow diagram if the efficiency,
$\epsilon$, with which $L_b$ is converted to radio emission is 
essentially constant. The normalization is also good if
$\epsilon \simeq 0.1$ (Gopal-Krishna \& Wiita, 2000b, in preparation).

\end{document}